\documentclass[journal=jpccck,manuscript=article]{achemso}

\usepackage{amssymb}
\usepackage{amsmath}
\usepackage{color}
\usepackage{subfigure}
\usepackage{graphicx}

\usepackage[version=3]{mhchem} 
\usepackage{natbib}
\usepackage{mciteplus}

\setkeys{acs}{usetitle = true}
\def\@dotsep{4.5}
\makeatother

\usepackage{bm}
\usepackage{graphicx}
\usepackage{amssymb}
\usepackage{amsmath}
\usepackage{hyperref}
\def \be {\begin{equation}\displaystyle}
\def \ee {\end{equation}}
\def \bea1 {\begin{eqnarray*}\displaystyle}
\def \eea1 {\end{eqnarray*}}

\newcommand{\BEA}{\begin{eqnarray}}
\newcommand{\BEAN}{\begin{eqnarray*}} 
\newcommand{\EEA}{\end{eqnarray}}
\newcommand{\EEAN}{\end{eqnarray*}}

\title{Graphdiyne Pores: 'Ad hoc' Openings for Helium Separation Applications}

\author{Massimiliano Bartolomei}\email{maxbart@iff.csic.es}
\affiliation{Instituto de F\'{\i}sica Fundamental,
Consejo Superior de Investigaciones Cient\'{\i}ficas (IFF-CSIC), Serrano 123,
28006 Madrid, Spain}

\author{Estela Carmona-Novillo} 
\affiliation{Instituto de F\'{\i}sica Fundamental,
Consejo Superior de Investigaciones Cient\'{\i}ficas (IFF-CSIC), Serrano 123,
28006 Madrid, Spain}

\author{Marta I. Hern\'{a}ndez}
\affiliation{Instituto de F\'{\i}sica Fundamental,
Consejo Superior de Investigaciones Cient\'{\i}ficas (IFF-CSIC), Serrano 123,
28006 Madrid, Spain}

\author{Jos\'e Campos-Mart\'{\i}nez}
\affiliation{Instituto de F\'{\i}sica Fundamental,
Consejo Superior de Investigaciones Cient\'{\i}ficas (IFF-CSIC), Serrano 123,
28006 Madrid, Spain}
\affiliation{Instituto de F\'{\i}sica Fundamental,
Consejo Superior de Investigaciones Cient\'{\i}ficas (IFF-CSIC), Serrano 123,
28006 Madrid, Spain}

\author{Fernando Pirani}
\affiliation{Dipartimento di Chimica, Biologia e Biotecnologie, Universit\`a
di Perugia, Perugia, Italia}

\author{Giacomo Giorgi}
\affiliation{Department of Chemical System Engineering, School of Engineering,
University of Tokyo, Tokyo, Japan}

\date{\today}

\begin{document}

\begin{abstract}
Two-dimensional (2D) materials deriving from graphene, such as graphdiyne and 2D
polyphenylene honeycomb (2DPPH), have been recently synthesized and 
exhibit uniformly distributed sub-nanometer pores, a feature 
that can be exploited for gas filtration applications.
Accurate first principles electronic structure calculations are reported
showing that graphdiyne pores permit an almost unimpeded helium transport while it is much more difficult through the 2DPPH openings.
Quantum dynamical simulations on reliable new force
fields are performed in order to assess the
graphdiyne capability for helium chemical and isotopic separation.
 Exceptionally high He/CH$_4$ selectivities are found in a wide
range of temperatures which largely exceed the
 performance of the best membranes used to date for helium extraction from natural
 gas. Moreover, due to slight differences
 in the tunneling probabilities of $^3$He and $^4$He, we also find promising
 results for the separation of the fermionic isotope at low temperature. 
\end{abstract}

\maketitle


\section{Introduction}

Membranes using nanoporous two-dimensional (2D) materials are emerging as
attractive candidates for applications in molecular separations and related
areas\cite{review2D:2013,review2D:2014}.
In particular, graphene-based materials\cite{Marlies:2013} exhibit vantage
points with respect to conventional 3D materials, by virtue of their thermal 
and chemical stability as well
as a low molecular weight and efficient transport capability.
Graphene sheets do not have pores for molecular sieving and the introduction
of sub-nanometer pores is required to obtain sufficient molecular
permeance. This can be achieved by means of ``top-down'' fabrication methods 
able to create holes in suspended graphene
sheets\cite{litography:2008}, which, however, are in general not sufficiently 
controllable.
``Bottom-up'' assembly processes represent a valuable alternative
and have permitted the synthesis of new 2D materials characterized
by regular and uniformly distributed sub-nanometer pores.  
The most promising examples are graphdiyne\cite{graphd-chemcomm:2010} and 2D 
polyphenylene honeycomb (2DPPH)
\cite{Bieri:09}. In the former, peculiar triangular pores arise from carbon 
chains
formed by two conjugated C-C triple bonds that link adjacent benzene
rings. In the latter, the resulting structure is instead similar to that of
graphene  but with one missing hexagon per unit cell which is passivated by 
hydrogens,
leading to almost circular openings (see Fig. 1.a).
 The successful synthesis of these materials has led to important theoretical
studies devoted, firstly, to their application as effective single-layer
membranes for gas separation and water filtration
technologies 
\cite{Bieri:10,Schrier:10,nanoscalemitbis:2012,nanoscalemit:2013,jpclours:2014}
and, secondly, to the tailoring of related porous 2D structures by changing
pore shape and size through the incorporation of functional groups
\cite{Hauser:2012,Schrier:13,nanoscalepph:2014,dft-vdw:14bis}.     

In this work we want to assess the capability of the recently synthesized
graphdiyne and 2DPPH materials as efficient membranes for helium
separation applications. 
As it is well known, helium is an irreplaceable natural resource and its 
growing demand in a variety of industrial
and scientific applications, together with ongoing  production deficiencies,
has recently led to shortages of this element\cite{Nature-He:2012}. 
Specifically, the lighter isotope, $^3$He, is crucial for large neutron-scattering facilities and the depletion of its stockpiling might also
affect fundamental research\cite{Nature-He:2014} in ultracold physics and chemistry.
Natural gas remains the richest and most accessible source of helium even if
its concentration is less than 1\% in most of the helium-producing wells. 
Unfortunately most of the natural-gas plants treat helium as a valueless gas
and vent it to the atmosphere. Therefore advanced and efficient technologies
for its ``in situ'' recovery are highly desirable.

Existing methods for performing both the chemical separation of He from
natural gas and the separation of He isotopes require energetically costly
techniques, such as cryogenic distillation and pressure-swing
adsorption\cite{He-adsorp:2008}.
In principle membrane-based gas separation has a much lower
thermodynamic cost\cite{membr-review:2009} and recent theoretical works have proposed the use
of 2D membranes for the chemical\cite{Schrier:10,Bieri:10,Schrier:13,silicene:13} and
isotopic\cite{Schrier:10,Hauser:2012,Schrier:12,ceotto:2014}  separation of He gas.  
 In particular, studies on He permeability through 2DPPH membranes were already
 addressed\cite{Schrier:10,Bieri:10} but, to our knowledge, the graphdiyne
 capability for He separation has not been investigated yet.
Moreover, in our opinion, the reliability of the He--2DPPH  penetration barriers,
obtained by means of the density functional theory (DFT)\cite{Bieri:10} and second-order
M{\o}ller-Plesset perturbation (MP2)\cite{Schrier:10} methodology, might need a deeper analysis. 

The paper is organized as follows.  Section 2 
refers the computational methodologies for electronic structure calculations
as well as the quantum mechanical tools used to calculate the penetration 
probabilities.   In Section 3 
we present our results concerning the barriers and
the quantum simulations using new force fields, optimized from the
electronic structure calculations. The paper ends with Section 4
, in which some conclusions are summarized.

\section{Computational Methods}
\label{sec.2}

The electronic structure calculations have been carried out at the ``coupled''
supermolecular second-order M{\o}ller-Plesset perturbation theory
(MP2C)\cite{mp2c} level of theory by using the Molpro2012.1 
package\cite{MOLPRO}.
The geometry of the systems is that of Fig. 1.a, and for the pores 
we have considered the following bond lengths\cite{mech-graphdiyne:2012}: 
1.431 \AA\, for the aromatic C-C, 1.231 \AA\,
for triple C-C, 1.337 \AA\, for the single C-C between two triple C-C bonds,  
1.395 \AA\, for the single C-C connecting aromatic and triple C-C bonds and 
 1.09 \AA\, for C-H bonds.
The aug-cc-pVTZ\cite{Dunning} basis set was employed for the pore structures, 
while the  aug-cc-pV5Z\cite{Dunning} basis has been used for rare gases 
and methane molecule. All considered molecular
structures are treated as rigid bodies: the atoms composing the
investigated pores are frozen in their initial positions and the
molecular configuration of methane is not allowed to relax during the calculations.  
The interaction energies have been further corrected for the basis set
superposition error by the counterpoise method of Boys and
Bernardi\cite{Boys:70}.

For the calculation of the transmission probabilities of the atom/molecule 
through the pores we have employed a model that makes use of wave packet 
propagation. 
This technique is specially well suited for this system\cite{hernandez-prb:94} 
since it is not necessary to invoke any periodicity and defects and many other
interesting features can be naturally taken into account.
The time-dependent Schr{\"o}dinger equation was solved by propagating wave
packets in a potential energy
surface that is symmetric with respect to the origin,  $V(-z)=V(z)$, where $z$
represents the distance between the atom and the membrane. The initial wave packet
is\cite{HellerGWP}  

\begin{eqnarray}
\psi(z,t=0) & = & \bigg( \frac{2 Im(\alpha)}{\pi \hbar}\bigg) ^{1/4} \nonumber \\
         & & \hspace{-0.75cm} \times \exp \Big\{ \frac{i}{\hbar} \Big[
\alpha (z-z_0)^{2} +   p (z-z_0) \Big] \Big\},
\end{eqnarray}

\noindent
where $\alpha=$ 0.434 $i$ a.u., $z_0$ is a sufficiently large distance
(typically 12-16 \AA), and $p$ is chosen such that $p^{2}/(2 \mu)$ is close to
  the barrier height, where $\mu$ is the mass of the colliding particle.
 The wave packet is represented in a grid from $z= -L$
to $z=L$ ($L$= 50 \AA) and is propagated using the split operator
method\cite{SplitOperator}. The transmission probability is obtained by
evaluating the probability current at the barrier ($z=0$)\cite{cplh2h2:01,Zhang:91}

\begin{equation}
P(E)=\frac{\hbar}{\mu} Im \bigg(  \psi^{\star}_E(z=0)   
\frac{\partial \psi_E }{\partial z}  \textbar_{z=0} \bigg),
\label{eq2}
\end{equation}

\noindent
 where $\psi_E$ is the stationary wave
function. It is computed from the Fourier transform of the time-dependent
wave packet, 

\be
 \psi_E (z) = \frac{1}{a(E)} 
\int dt \; e^{iEt/\hbar} \; \psi(z,t),
\ee

\noindent
with

\be
a(E) = C \int dz e^{i k z} \psi (z,t=0)
\ee

\noindent
where $k= \sqrt{2 \mu E}/\hbar$ and $C= \sqrt{\mu/(\hbar k)}$.  
It is important to note
that, with this choice for the normalization of the stationary wave function,
the probability current of Eq. (2)  becomes adimensional and, in
addition, it exactly corresponds to the transmission 
probability\cite{Zhang:91,Miller:74}.

 Finally, the temperature-dependent
transmission probability, $P(T)$, was computed from numerical integration of
$P(E)$, weighted by a one-dimensional Maxwell-Boltzmann distribution of
molecular velocities.

\section{Results and Discussion}
 
\subsection{Penetration Barriers}

In Fig. 1.a we report the molecular structures that can be
considered as the smallest precursors\cite{Schrier:10,jpclours:2014} of
graphdiyine and 2DPPH and which are here used to study their pores.
It can be seen that, inside the pores, smaller black figures are also
depicted. Their area corresponds to the effective pore size which has
been estimated by considering the van der Waals (vdW) radii of the acetylenic
carbon bonds, for graphdiyne, and of the inner hydrogen atoms, for 2DPPH ($\sim$ 1.6 \AA\,
and 1.3 \AA, respectively). The side length of the inner triangle is  3.9 \AA, for graphdiyne, and the diameter of the
inner circle is 0.8 \AA\, for 2DPPH. 
In the same way, the helium vdW diameter is 2.6 \AA\cite{HeHe:Tang}(represented by a double-headed arrow in
 Fig. 1.a) and it is smaller than the graphdiyne pore side but larger than
the diameter of the 2DPPH circular opening.
This comparison already suggests that He penetration should be
easy in graphdiyne pores and more difficult when passing through the
2DPPH smaller pore. 

A more quantitative assessment comes from electronic structure calculations.
The prototypical systems of Fig. 1 are small enough to 
allow the use of the high level (and computationally expensive) MP2C
approach and, at the same time,  they are large enough to describe
  the interaction features of the pores, as demonstrated by test calculations
  in which we have considered further prototypes of increasing size. 
The accuracy of the MP2C method, which is particularly suited to recover
non-covalent weak interactions, has been recently assessed by
extensive calculations of rare gas--fullerene\cite{mp2c-full} and  -coronene\cite{grapheneours:2013}
interaction energies and it was concluded that the MP2C results are 
in good agreement with density functional theory-symmetry
adapted perturbation theory (DFT-SAPT)\cite{dft-sapt} calculations, with the
advantage of a significantly lower computational cost.
In Fig. 1.b we report the calculated potential energy profiles for one
helium atom perpendicularly approaching the 2D structures. In particular, $z$ is the distance between helium
and the geometric center of the pores.    
It can be seen that the penetration barrier (defined as the
energy difference between $z$=0 and $\infty$, E$_{PB}$) is high (about 0.516
eV) for 2DPPH but becomes much lower (about 0.033 eV) for
graphdiyne. 

The He--2DPPH penetration barrier was previously obtained by means of MP2 and DFT calculations and found to be
0.523\cite{Schrier:10} and 0.430\cite{Bieri:10} eV, respectively. 
Our estimation is in between the previous calculations and closer to the
MP2 value. Present value should be considered as the
reference one since, to our knowledge, the employed
dispersion corrected DFT approach\cite{Bieri:10} has not been sufficiently
tested on this specific system. In fact we have determined that the calculated
  penetration barrier may vary depending on both the adopted density
  functional and 'a posteriori' empirical dispersion correction.
 Also, in our opinion the MP2 estimation\cite{Schrier:10} has been obtained by using basis sets (cc-pVDZ and
   cc-pVTZ) not sufficiently extended, thus neglecting
   diffuse functions that we have found to be important to properly describe
   the interaction in this system.
Moreover the standard  MP2 approach lacks the
correction included in the MP2C method\cite{mp2c} which is indeed needed
to improve the description of non-covalent interactions\cite{mp2c-full}.
  As an example, we have estimated that by using the same basis set the penetration
barrier at the MP2 level of theory would be about 7\% larger than present  MP2C
  value.
As for the He--graphdiyne system, it should be stressed that the calculated
barrier of few tens of meV is not only much lower than that for 2DPPH, but also
represents an energy gap comparable with other proposed 2D
materials which, however, have not been fully synthesized yet. They include partially nitrogen functionalized porous
graphene\cite{Hauser:2012} (about 0.025 eV) and  porous graphene-E-stilbene-1 \cite{Schrier:13}
(about 0.050 eV).

Barrier heights could be further reduced if the optimization of the
position of the atoms defining the pores is taken into account. In fact, related  additional
  calculations at the density functional level of theory (DFT) have been
  performed by considering the full ``periodic'' structure of a single
  graphdiyne layer and we have found that the barrier can lower by about 15$\%$ due to
  the deformation of the pores (see Fig. S1 of the Supporting Information).

To examine the efficiency of the two kinds of pores for helium
permeation in more detail, we have considered a simple Arrhenius behavior for the diffusion
rate $D$ as a function of the temperature $T$: $D(T)= A_0 e^{\frac{-E_{PB}}{kT}}$,
where $E_{PB}$ is the computed penetration barrier. 
As it is custumary in this context\cite{Bieri:10,silicene:13,nanoscalepph:2014} we have taken the same diffusion prefactor $A_0$=10$^{11}$s$^{-1}$ for both pores.
The related diffusion rates are reported in Fig. 1.c and it can be seen that
graphdiyne exhibits a much higher permeability than 2DPPH in an ample range of temperature, being the difference
of about eight orders of magnitude at room temperature. 
These results clearly show that, due to its ``ad hoc'' pore size, a
graphdiyne membrane is much more suited than the 2DPPH one for helium permeation and
in the following we will focus on the former, analyzing in detail its separation capabilities.   

In order to investigate the separation of helium from natural gas,
 we have considered neon and methane as limiting cases of the complex
 mixture of alkanes and rare gases therein usually present. Thus, additional accurate
 energy profiles are reported in Fig. 2 (upper panel) and compared with
 that of helium.
It should be pointed out that methane is considered as a pseudo-atom, so that
 the reported energy profile is actually an average of different potential
 curves(see Supp. Inf. and Fig. S2). We find that this approximation is
 reasonable since the collision time $t_c$ (the average time taken to cross
 the barrier) is about 2.5 times larger than the average CH$_4$ rotational
 period, $t_r$, for temperatures ranging from 100 to 300
 K\cite{LevineBersteinBook,Note-tctrot}.
We have found that the penetration
barrier $E_{PB}$ (also reported in Table 1) for Ne and CH$_4$ increases up to 0.106 and
1.460 eV,  respectively, suggesting a high impediment to the passage of methane.

\subsection{Force Fields and Quantum Dynamical Simulations}

To assess the separation capability of the graphdiyne pores, we have computed
transmission probabilities for the passage of these species by means of time-dependent wave packet simulations.
A new force field, optimized on the benchmark MP2C electronic
  structure results, 
 has been obtained to this end.  It provides the basic features of the
 interaction in the full configuration space, and therefore it is suitable for
 performing molecular dynamics simulations. Specifically, for describing the non-covalent
 interaction between the rare gas (and methane in the pseudo-atom limit) and the
carbon atom forming the graphdiyne net structure the Improved
Lennard-Jones (ILJ) pair potential function\cite{ILJ} is used: 

\begin{equation}
V_{rg,c}(R)=
\varepsilon\left[\frac{6}{n(x)-6}\left (
\frac{1}{x} \right)^{n(x)} -
\frac{n(x)}{n(x)-6}\left(
\frac{1}{x}\right)^6 \right]
\label{eq:1}
\end{equation}

\noindent
where $x$ is a reduced pair distance $x = \frac{R}{R_{m}}$, and  $\varepsilon$
and $R_{m}$ represent the well depth and equilibrium distance of the
rare gas(methane)-carbon interaction, respectively.
Moreover,  $n(x)$ is expressed by \cite{Pirani:04} $n(x)= \beta + 4.0 \; x^2$
where $\beta$ is a parameter defining the shape of the potential and depending
on the nature and hardness of the interacting partners.
The optimized ILJ parameters are reported in Table 1.
They have a physical meaning, since they have been obtained
by fine tuning initial data estimated by exploiting the polarizability values
of the interacting partners. In the case of the graphdiyne pore, the average
effective polarizability of the C atom has been used.
 In particular, a value equal to 1.1\AA$^3$, consistent with that
  reported by Gavezzotti\cite{gave:03} describing
  the average behavior of C atom in various aliphatic and unsaturated
  molecules, has been adopted to estimate the dispersion energy contribution
  and the $\varepsilon$ parameter associated to each rare gas (methane)-C
  interacting pair (see also Ref. \cite{margarita:12tris}). Moreover, many-body
  effects have been also taken into account in the evaluation of the $R_m$ values according
  to Ref. \cite{Pirani:01}.

 Additional MP2C energy curves for approaches of the atom (molecule)
 to different sites of the graphdiyne pore have been computed 
 and taken into account (see Fig. 3) in order to more extensively test the features of the involved force field.
In the upper panel of 
Fig. 2 (and in Fig. 3) we present a comparison between MP2C and ILJ curves and a very good agreement can be observed
which confirms the reliability of the proposed force fields. 

For simplicity we have  just considered a one-dimensional 
transmission\cite{Schrier:10,Hauser:2012} of the atom through the center of a 
graphdiyne pore, where the interaction potentials have been obtained by
summing up ILJ pair potentials between the atom (molecule) and the 
neighboring carbon atoms of a graphdiyne sheet until convergence. 
 Values of the penetration barriers obtained in this way are reported in Table 1. 
Thermally weighted transmission probabilities $P(T)$ have been computed and
used to estimate selectivities of the different molecular
  combinations $X/Y$, defined as the probability ratios
$S=P^X/P^Y$. The temperature dependence of these selectivities is
  presented in the lower panel of Fig. 2. Exceptionally high selectivities can be
noticed for He/CH$_4$ and Ne/CH$_4$ combinations in a wide range of
temperatures and they are about 10$^{24}$ and 10$^{23}$, respectively, at
room temperature.  As for the He/Ne combination, the corresponding
selectivity is about 27 at room temperature, suggesting a less efficient
capability for the separation of these species, but in any case larger than 6,
the value considered acceptable for industrial applications\cite{permeance:06}.
It should be noticed that the predicted selectivities could be actually
  reduced if the lowering of the penetration barrier due to deformation of the
  pores is taken into account: for the heavier species the
  larger lowering is expected (see Fig. S1) even if we do not consider that this effect
  can alter significantly the very favourable He/CH$_4$ and Ne/CH$_4$ probability ratios 
 we have found.
 Further effects due to the neglect of both the interaction and
competition of single components in the gas mixtures are not expected to
provide significant deviations\cite{Schrier:13} in the obtained $S$
behavior at least around and above room temperature.

Our results clearly indicate that the combination of a very high He
diffusion rate (Fig. 1) and very high He/CH$_4$ and Ne/CH$_4$ selectivities
(Fig. 2) makes graphdiyne a very promising material for the separation of He
and Ne from the hydrocarbons contained in natural gas (higher alkanes,
 also present in natural gas, are expected to have even smaller transmission
probabilities due to their larger size). As a matter of fact, due to the very low He and Ne
concentrations in most of the natural gas reservoirs, very high selectivities
(about 1000\cite{Membrane-handbook:2001}) are required.
 This stringent requirement is indeed largely overcome by present estimations
 and, to our knowledge, not achieved by the polymeric
 membranes proposed up to date (corresponding selectivities are about
 100\cite{heliumcsic:2014}).

The high permeability that we have found for He-graphdiyne has 
encouraged us to study the possibility of separation of its isotopic
variants. In particular, the separation of $^3$He and $^4$He could be achieved
by taking advantage of slight differences in the tunneling probabilities of the
two isotopes and by exploiting them in multistage processes\cite{ceotto:2014},
which involve the 
passage of the gas through a number of graphdiyne layers. To this end we have
computed transmission probabilities for $^3$He and $^4$He as functions of
kinetic energy. 
The results are given in the upper panel of
Fig. 4. It can be seen that, for kinetic energies lower than the classical barrier
(see last column of Table 1), the $^3$He transmission probability is
significantly higher than that of 
$^4$He, while the opposite occurs for higher kinetic energies. For this
reason, it makes sense to keep the gas temperature as low as possible in order
to exploit the low energy side of the transmission curve.

Thermally weighted transmission
probabilities as well as the corresponding $^3$He/$^4$He
selectivity 
as a function of temperature are reported in
Fig. 4 (lower panel).
It should be stressed that, for simplicity, we have assumed a (classical) Boltzmann distribution for kinetic energies of both isotopes instead of the intrinsic quantum statistics of fermionic $^3$He and bosonic $^4$He atoms. However, if a mixed nonideal quantum gas was considered, an even larger $^3$He/$^4$He selectivity would be expected since the Fermi-Dirac distribution is broader than the  Bose-Einstein one.
Therefore, present estimation for the isotopic selectivity at 77 K (liquid nitrogen temperature) is about 1.04 and
increases rapidly at low
temperatures reaching the acceptable reference value of
6\cite{permeance:06} at about 20 K.
However, as the temperature drops, the transmission probabilities considerably decrease, as well as the He flux. 
In fact at 20K the $^3$He transmission probability, even if larger than that of
$^4$He and of the classical limit (see lower panel of Fig. 4), is quite low and in the range of 10$^{-9}$\cite{Note-barrier}. 
The overall flux is the
transmission probability  multiplied by the frequency of gas collisions
with the pore, which, according to the kinetic theory of gases, is
$P/\sqrt{2\pi \mu kT }$
 where $P$ is the pressure of the species, $T$ 
 the temperature and $k$ the Boltzmann constant.
 If we assume a 100$\%$ porous sheet, ideal gas conditions and a pressure of 3
 bar we can estimate an upper-bound\cite{Note-bound} of the total helium flux:  at 77K it is about 3.7$\cdot$10$^{-3}$ moles cm$^{-2}$
    s$^{-1}$  while at 20K is about 1.5$\cdot$10$^{-8}$ moles cm$^{-2}$
     s$^{-1}$, a value  leading to a permeance\cite{Note-permeance} equal to 1.5$\cdot$10$^{-8}$ moles cm$^{-2}$
     s$^{-1}$ bar $^{-1}$ which is slightly lower than the limit of 6.7$\cdot$10$^{-8}$ moles cm$^{-2}$
     s$^{-1}$ bar $^{-1}$\cite{permeance:06} considered acceptable for
     industrial applications.

\section{Conclusions}

In summary, by means of electronic structure MP2C computations,
 we have shown that graphdiyne is much more suited than 2DPPH for helium
permeation since it involves a much lower penetration barrier (of the order of
few tens of meV). Additional calculations of the interaction of
 neon and methane with a graphdiyne pore have allowed us to setup reliable full force fields which
 have been used to compute quantum mechanical probabilities and selectivities
 for the passage of gases through the openings. We have found exceptionally favorable
 He/CH$_4$ and Ne/CH$_4$ selectivities in a wide range of temperatures
 which largely exceed the performance of the best membranes used for helium
 chemical separation. Thermally weighted helium tunneling calculations have
 been also performed to show selective transmissions of $^3$He versus $^4$He.
 For temperatures below 100K the $^3$He/$^4$He selectivity becomes appreciably
 larger than 1, reaching an acceptable value of 6 at about 20K.
 We thus propose that, in addition to its promise
for hydrogen\cite{nanoscalemitbis:2012} and water\cite{jpclours:2014} purification, a graphdiyne based
 membrane can be efficiently used for helium separation applications addressed to filtering natural gas and, possibly,
 isotopic mixtures. 
On the basis of present results, since species other than helium are assumed to provide
penetration barriers whose values increase with their vdW diameter,
we consider that graphdiyne pores should also provide
 favorable selectivities for He/Ar, He/Kr and He/N$_2$ gas combinations
 which could  permit optimal separation of helium from heavier rare gases
 and hydrothermal spring gases, the latter primarily composed of nitrogen\cite{hydrothermal:03}.

A full assessment of the capability of graphdiyne for helium isotopic separation would
actually require three-dimensional simulations in order to better
estimate the helium permeance at low temperatures and to take into account for
zero-point energy effects due to in-pore vibrational modes.
Work in this direction, by using the ILJ full force field, is in progress.

\section*{Acknowledgments}

The work has been funded by Spanish grants FIS2010-22064-C02-02 and FIS2013-48275-C2-1-P. 
Allocation of computing time by CESGA (Spain) 
is also  acknowledged.
F.P. acknowledges financial support from the Italian Ministry of University 
and Research (MIUR) for PRIN 2010-2011, grant 2010 ERFKXL\textunderscore002.

\section*{Supporting Information}
  Periodic DFT calculations as well as
  intermolecular potentials related to different configurations
  used for graphdiyne pore--methane computations are reported in
  additional figures.
  This information is available free of charge via the Internet at http://pubs.acs.org.


\begin{mcitethebibliography}{52}
\bibitem[Not()]{Note-tctrot}
The collision time was estimated as $t_c= L/\langle v \rangle$, L being the
  range of the interaction (specifically, the distance from the van der Waals
  minimum to the barrier), and $\langle v \rangle$, the mean speed at a given
  temperature. Methane rotational time was computed as $t_r= 3 \pi \hbar / [B
  \sqrt{J^{\star}(J^{\star}+1)}]$, where $B$ is the rotational constant,
  $J^{\star}$ is the most probable angular momentum at a given temperature, and
  the factor three accounts for the symmetry of the molecule\relax
\mciteBstWouldAddEndPuncttrue
\mciteSetBstMidEndSepPunct{\mcitedefaultmidpunct}
{\mcitedefaultendpunct}{\mcitedefaultseppunct}\relax
\EndOfBibitem
\bibitem[Not()]{Note-barrier}
It should be noticed that, due to the pore relaxation, the reduction of the He
  penetration barrier can also lead to a higher $^3$He trasmission probability
  which at 20K would be about 2$\cdot 10^{-8}$. The corresponding permeance
  would also increase to 2.4$\cdot$10$^{-7}$ moles cm$^{-2}$ s$^{-1}$ bar
  $^{-1}$ leading to a better agreement with the acceptable value for
  industrial applications.\relax
\mciteBstWouldAddEndPunctfalse
\mciteSetBstMidEndSepPunct{\mcitedefaultmidpunct}
{}{\mcitedefaultseppunct}\relax
\EndOfBibitem
\bibitem[Not()]{Note-bound}
We assume that any particle reaching the surface will always encounter a pore
  and that it will experience a potential energy barrier equals to that normal
  to the plane of the pore. In a more realistic simulation the helium flux is
  expected to be lower.\relax
\mciteBstWouldAddEndPunctfalse
\mciteSetBstMidEndSepPunct{\mcitedefaultmidpunct}
{}{\mcitedefaultseppunct}\relax
\EndOfBibitem
\bibitem[Not()]{Note-permeance}
Permeance ($Q$) is defined as the ratio between the flux and pressure drop
  across the pore, $Q=Flux/\Delta P$. In the present case we are assuming
  $\Delta P$= 1 bar.\relax
\mciteBstWouldAddEndPunctfalse
\mciteSetBstMidEndSepPunct{\mcitedefaultmidpunct}
{}{\mcitedefaultseppunct}\relax
\EndOfBibitem
\end{mcitethebibliography}

\providecommand*\mcitethebibliography{\thebibliography}
\csname @ifundefined\endcsname{endmcitethebibliography}
  {\let\endmcitethebibliography\endthebibliography}{}


\begin{figure}[h]
\includegraphics[width=7.5cm,angle=0.]{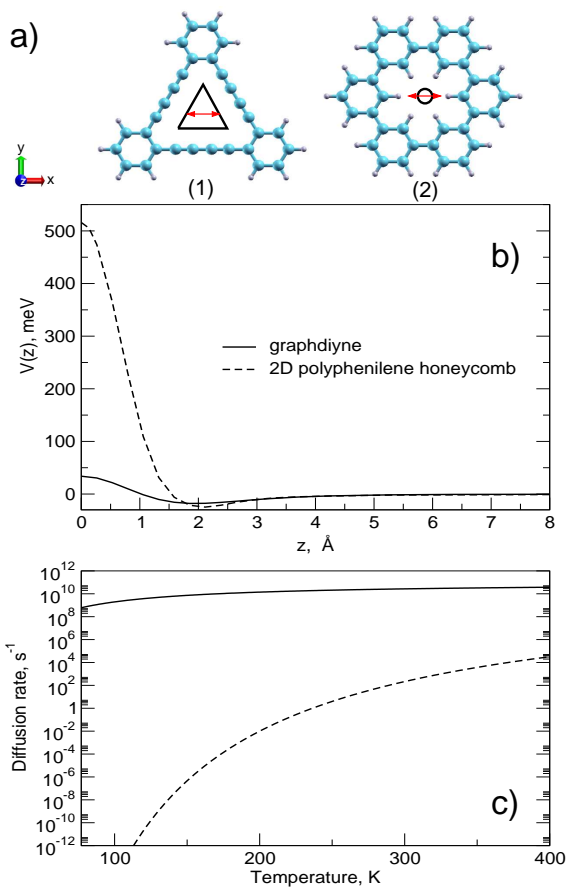}
\caption[]{a) Molecular structures used to study the nano-pores of
 graphdiyne(1) and 2D polyphenylene honeycomb (2).
The black triangle and circle depicted inside the pores represent their effective available area  
to be compared with the van der Waals diameter of the He atom (red double-headed arrow).
b) Energy profiles  obtained at the MP2C level of theory  for He
perpendicularly  approaching the geometric center of graphdiyne and 2DPPH pores.
c) Rates for He diffusion through the pores as functions of temperature (see text).}
\label{fig1}
\end{figure}

\begin{figure}[t]
\includegraphics[width=7.5cm,angle=0.]{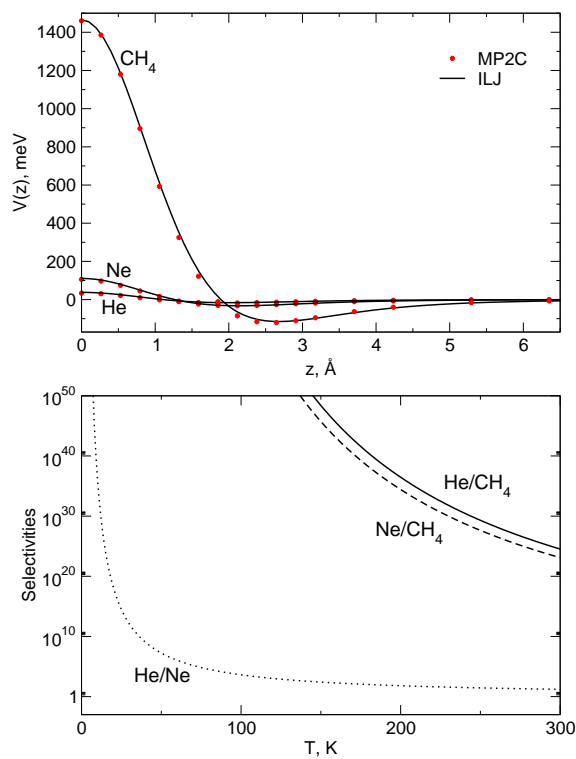}
\caption[]{Upper panel:
Energy profiles obtained for helium, neon and methane
perpendicularly  approaching the geometric center of the graphdiyne pore. In
the case of methane the resulting curve is the average of four limiting
configurations (see Fig. S2). Red dots refer to the MP2C level of theory results
while solid curves represent estimations obtained with optimized Improved
Lennard Jones (ILJ) force fields (see text).
Lower panel:
Selectivities for different molecular combinations  as functions of
temperature(see text).}
\label{fig2}
\end{figure}

\begin{figure}[t]
\includegraphics[width=7.5cm,angle=0.]{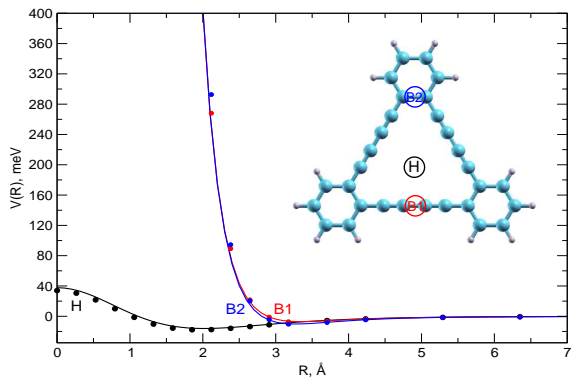}
\caption[]{
Interaction profiles for the approach of the helium atom to three
different sites (H, B1 and B2) of the graphdiyne pore.
Circles correspond to the MP2C {\it ab initio} results while solid lines
represent ILJ full force field predictions.}
\label{fig3}
\end{figure}

\begin{figure}[t]
\includegraphics[width=7.5cm,angle=0.]{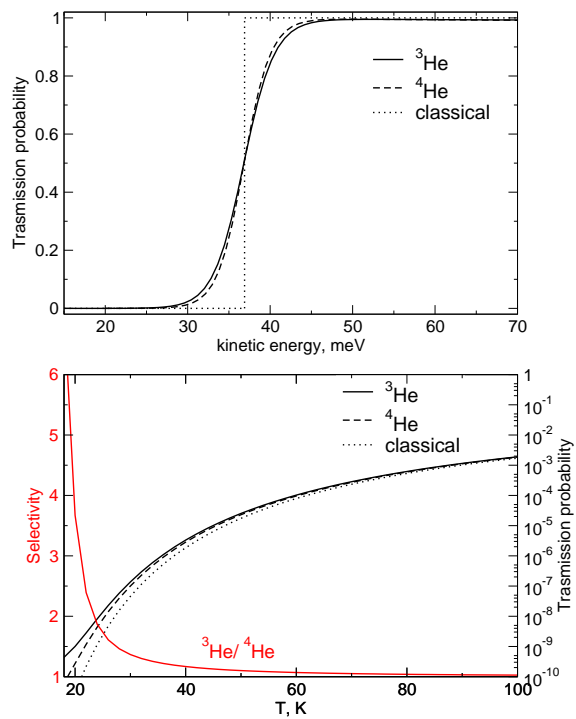}
\caption[]{Upper panel:
Quantum mechanical and classical transmission probabilities of He through
graphdiyne pore as a function of kinetic energy.
Lower panel:
$^3$He/$^4$He selectivity (red line) and thermally weighted quantum transmission probabilities (solid and dashed black
lines for $^3$He and $^4$He, respectively) as functions of
temperatures. Classical transmission probability is reported as a dotted line.}
\label{fig4}
\end{figure}  

\newpage

\begin{table}
\small
\caption{Penetration barrier(E$_{PB}$)  for He, Ne and
  CH$_4$   passage through  graphdiyne pores. 
  Parameters for the rare gas (methane)-carbon Improved
  Lennard Jones (ILJ) pair potential (see Eq. 5),
 used to obtain the full force fields, are  also reported, together with the
 resulting penetration barriers after a pair-wise summation over a graphdiyne
 sheet. E$_{PB}$ and $\varepsilon$ are in meV, R$_{m}$ in \AA~, and $\beta$ is
dimensionless.
}
\label{tab:1}
\begin{tabular}{cccccc}         
   & & \multicolumn{4}{c} {ILJ force field} \\
\cline{3-6} \\
       &  E$_{PB}(MP2C)$   & R$_m$ & $\varepsilon$ & $\beta$  & E$_{PB}(ILJ)$     \\
\hline
  & & & & &   \\     
   He &   33.90   &  3.595 & 1.209 & 7.5   &  36.92     \\
   Ne &  105.76    &  3.671 & 2.388  & 7.5   & 109.17   \\ 
  CH$_4$ & 1460.31   &  4.046 & 7.763 &  7.5  & 1454.70   \\ 
  & & & & &   \\ 
\hline
\end{tabular}
\end{table}

\clearpage
\newpage

\begin{large}
{\bf Table of Content}
\end{large}

\vspace{1.5cm} 

{\bf Graphdiyne is a novel two-dimensional material deriving from graphene
that has been recently synthesized and featuring uniformly distributed
sub-nanometer pores.}
Accurate calculations are reported showing that graphdiyne pores permit an almost unimpeded
helium transport which can be employed for its chemical and isotopic
separation. Exceptionally high He/CH$_4$ selectivities are found which largely exceed the
 performance of the best membranes used to date for extraction from natural
 gas. Moreover, by exploiting slight differences
 in the tunneling probabilities of $^3$He and $^4$He, we also find promising
 results for the separation of the fermionic isotope at low temperature. 

\vspace{2.5cm} 

\begin{figure*}
\includegraphics[width=12.cm,angle=0.]{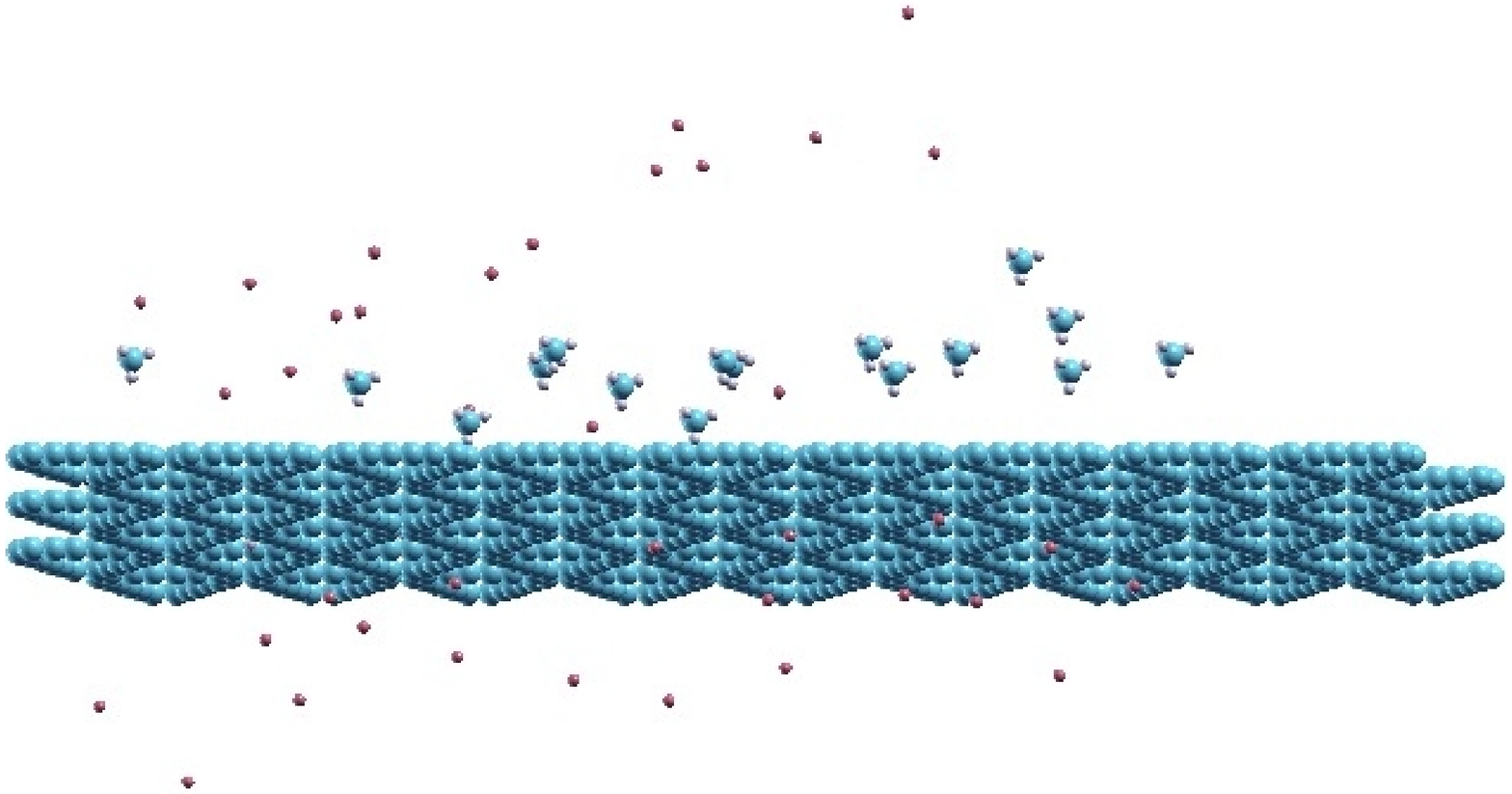}
\end{figure*}

\end{document}